\documentstyle[12pt]{article}

\begin{document}
\title{Transitions in Spectral Statistics}
\author{C. Blecken$^1$, Y. Chen$^2$ and K. A. Muttalib$^{1
\dag,2}$\\
$^1$Department of Physics, University of Florida\\
Gainesville, FL 32611, U. S. A.\\
$^2$Department of Mathematics\\
Imperial College, London SW7 2BZ, U. K.}
\maketitle
\begin{abstract}
We present long range statistical properties of a recently introduced
unitary random matrix ensemble, whose short range correlations were
found to describe a transition from Wigner to Poisson type as a function
of a single parameter. We argue, by evaluating the two-level correlation
function of a different solvable model, that the transition is perhaps a
quite general feature for a
class of models. Our analytic results compare well
with numerical studies on a variety of
physical systems. We discuss the possibility of experimentally
observing
signature of such transitions.
\end{abstract}
\newpage
Statistical properties of eigenvalues of matrices describing a wide
variety of quantum systems follow the universal results of random
matrix models \cite{LesHouches}. Such models were originally conceived
by Wigner to provide theoretical framework for the
understanding of the statistics of energy levels of heavy nuclei
\cite{Porter}, and has been developed into an important branch of
mathematical physics through the work of Dyson and Mehta \cite{Mehta}.
On very general grounds the randomness of the matrix elements, subject
to any relevant symmetry requirements, gives rise to a model of
eigenvalues repelling each other with a logarithmic interaction,
resulting in strong correlations amongst them. The normalizability
condition of the joint probability distribution requires a confining
potential for the eigenvalues, which can be thought of as resulting
from some physical constraint (e.g. a given eigenvalue density)
\cite{Balian}. For a given symmetry of the matrix, as long as
the eigenvalues are confined well, the statistical properties of the
levels in the bulk of the spectrum seem to be independent of the
particular choice of the constraint \cite{Nagao}, and follow universal
distribution known generally as the Wigner distribution.
However, it
is becoming increasingly evident that while the level statistics of a
wide variety of systems can be described very well by the highly
correlated Wigner distribution, many of these systems show a transition
to a completely uncorrelated Poisson distribution when some relevant
parameter is changed \cite{LesHouches}. Such transitions in the
spectral statistics might correspond to e.g. a chaotic-regular or a
metal-insulator transition in the system. Attempts have been made to
describe the transition in one particular statistical property, namely
the nearest neighbor spacing distribution (which is sensitive to only
the short range correlations between eigenvalues), either by purely
heuristic interpolation schemes \cite{Brody} or by considering
intermediate regimes where the phase space is partly chaotic and
partly regular \cite{Berry}. These results differ qualitatively from the
one case where the transition has been studied in detail numerically,
namely the case of metal to insulator transition in a disordered
system described by the microscopic random tight binding Anderson
Hamiltonian \cite{Shlovski}. As far as we know, no attempt has been
made to explain even heuristically the transition in other
statistical properties, such as the number variance, $V_n,$ or the so
called
$\Delta_3$ statistics which provides a quantitative
measure of the long range rigidity of the spectrum,
although there exist numerical evidences
for such transitions in both chaotic \cite{Jose} and disorderd systems
\cite{Muttalib2,Schrieber}.
\par\indent
Recently we have argued that the appropriate random transfer matrix
model related to disordered conductors belong to a new family of
random matrices; a solvable model then predicted a very specific type
of transition in the nearest neighbor spacing distribution as a
function of a single parameter \cite{Muttalib}. In the present work we
first show, by evaluating the two-level correlation function of a very
different (though related) but still solvable model, that the nature
of this transition in spectral statistics is not peculiar to the model
considered in \cite{Muttalib}, but is perhaps generic for a class
of models with weak confining potentials. We then calculate, for the
above models, two other statistical properties which describe the
transition in the long range correlation in the bulk of the spectrum,
namely the number variance and the $\Delta_3$ statistics. We show that
the variance of a linear statistic is controlled by a single parameter
in these models, and
is no longer universal. Our result includes the theorem of Dyson and
Mehta \cite{Mehta} on the universality of the variance of any linear
statistic as a special case in the appropriate limit. As an example, we
give explicit expression
for the number variance. The nature of the transition in
the $\Delta_3$ statistics agrees with earlier numerical results for
transfer matrices in disordered systems \cite{Muttalib2}.
\par\indent
It is well known that the  Wigner distribution is satisfied in the
appropriate range by eigenvalues of transfer matrices, Hamiltonians as
well as evolution operators \cite{LesHouches}. Same is true at the
other end of the transition, the Poisson distribution. Such
universality in the statistics results from the fact that once the
symmetries of the matrix have been properly taken into account, most
of the differences in
the detail appear in the local density of states, which disappear when
the distributions are described in units of the average local spacing.
It is therefore quite possible that the transition from Wigner to
Poisson distributions also has some broad general characteristics,
although there may be more than one class of transition depending on
the physical nature of the problem. It is therefore of interest to
compare the predictions of the above mentioned solvable model with
numerical results for different statistical properties of different
systems, described by different kinds of matrices. We find that the
transition in the $\Delta_3$ statistics describes
qualitatively well the numerically obtained statistics for a
variety of physical systems \cite{Muttalib2,Jose}, including the energy
levels of disordered
systems \cite{Schrieber}. Moreover, the transition in the
eigenvalue-spacing
distribution evaluated earlier for the model also shows remarkable
similarity to the transition in the distribution of energy levels
numerically obtained in ref. \cite{Shlovski} mentioned above.
\par\indent
We therefore conclude that in particular, the numerical results for
both the nearest neighbor spacing distribution and the long range
$\Delta_3$ statistics of the energy levels of disordered systems are
well described by the solvable models, even though the models were
originally constructed for transfer matrices.
The solvable models then allow us to calculate and predict further
consequences of such transitions in the energy level statistics. In
particular we predict how the ``correlation hole'' in the Fourier
transform of the absorption spectrum of small metallic particles
should be destroyed with increasing disorder, and discuss the
possibility of observing it experimentally.
\par\indent
As mentioned earlier, the random matrix models are characterized by
the confining potential $V(x)$. The statistical properties of the
levels can then be evaluated from the two level correlation function
which can be obtained explicitly from a set of orthogonal polynomials
defined with the potential as the weight factor \cite{Mehta}.
We will call the random matrix ensemble introduced in ref.
\cite{Muttalib} the ``$q-$Hermite'' Unitary Ensemble, because the
orthogonal polynomials defined by the potential are the
``q-generalization'' \cite{Askey} of the classical Hermite polynomials
that
characterize the conventional Wigner or Gaussian Unitary Ensemble
(GUE). The correlation function for the $q-$Hermite ensemble was found
to depend crucially on some parameter $\beta=\ln(1/q)$ characterizing the
potential:
\begin{equation}
K(u,v)={\beta\over 2\pi}{{\rm sin}\pi (u-v)\over {\rm
sinh}(\beta(u-v)/2)},
\end{equation}
where the scaled variables $u$ and $v$ are such that the density
$K(u,u)$ is unity.
As noted in ref. \cite{Muttalib}, this reduces to the corresponding
correlation function for the GUE in the limit $\beta\rightarrow 0$.
The above expression for the kernel is valid in the bulk of
the spectrum and for $0\leq\beta<2\pi^2.$ We
observe here that the nearest neighbor spacing distribution as a
function of the parameter $\beta$ obtained earlier for this model is
remarkably similar to the one obtained numerically in \cite{Shlovski}
for the energy eigenvalues corresponding to a microscopic Anderson
model for disordered conductors going through a metal-insulator
transition. (The difference in the power law behavior for small
spacing and the precise point where all different curves cross is
due entirely to the fact that the numerical results are for orthogonal
symmetry, while our model has unitary symmetry.) In order to establish
that the above two-level kernel is not a peculiarity of the $q-$Hermite
ensemble, we first consider another ensemble defined by the potential,
\begin{equation}
V(x;q)=\sum_{n=0}^{\infty}\ln[1+(1-q)q^nx],
\end{equation}
where the eigenvalue $x$ is from $0$ to $\infty$ as opposed to the
range $-\infty$ to $+\infty$ for the $q-$Hermite ensemble. For large $x$
and small $q$, this potential behaves as $[\ln x]^2$, as in the
$q-$Hermite case, but for small $x$ it behaves linearly as opposed to
the quadratic dependence in $q-$Hermite. This potential was considered
in \cite{Chen} as a possible model for the transfer matrix describing
disordered conductors. The orthogonal polynomials for this potential
are the $q-$Laguerre polynomials, a generalization of the classical
Laguerre polynomials \cite{Hahn}; we therefore call it the
``$q-$Laguerre'' model. From the asymptotic properties of the
$q-$Laguerre polynomials \cite{Chen}, we obtain the two-level kernel
in the bulk of the spectrum as a function of $\beta=\ln(1/q)\gg 1$ in the
limit where the number of eigenvalues $N\rightarrow \infty$:
$${\bar K}(x,y)$$
\begin{equation}
\;={{\rm cst}\over x-y}\;
\left[\left(\frac{x}{y}\right)^{1/4}{\rm sin}\left(\frac{\pi}{2\beta}
{\rm ln}x\right) {\rm cos}\left(\frac{\pi}{2\beta}{\rm ln}y\right)-
\left(\frac{y}{x}\right)^{1/4}{\rm sin}\left(\frac{\pi}{2\beta}{\rm
ln}y\right) {\rm cos}\left(\frac{\pi}{2\beta}{\rm ln}x\right)\right].
\end{equation}
In order to compare with the Wigner distribution for which the average
spacing between adjacent levels is unity, it is necessary to use a
transformation of variables $x={\rm e}^{2\beta u},\, y={\rm e}^{2\beta
v}$, such that the density in the new variable is uniform and unity.
In this new variable, the kernel becomes precisely the same as given
in Eq.(1) for $q-$ Hermite ensembles. It therefore
follows that the statistical properties of the levels in the two
models are identical in the bulk. Note that for $\beta\rightarrow 0$,
the two models reduce to classical Hermite and Laguerre models, which
are known to have identical kernels in the bulk \cite{Nagao}. (We
shall not consider here the interesting ``edge effects'' in the
Laguerre ensemble
\cite{Stone}.) Thus the statistical properties of the levels in the
bulk of the spectrum for large $N$ are insensitive to the details of
the model as long as some general features of the model (weak
$[\ln x]^2$ confinement of the large eigenvalues) remain the same.
\par\indent
We now evaluate the number variance and the $\Delta_3$ statistics for
the $q-$ Hermite model, expecting that these results will be valid for
at
least a class of similar models. The variance $V_n={\overline
{n^2}}-{\overline n}^2$ of the number of eigenvalues $n$ in an
interval $(-s/2,s/2)$ can be expressed as \cite{Mehta}
\begin{equation}
V_n(s)=\int_{-s/2}^{s/2}dx\int_{-s/2}^{s/2}dy\left[\delta(x-y)-Y(x-y)
\right]={2\over \pi}\int_{0}^{\infty}dk{1-{\rm cos}ks\over k^2}
\left[1-b(k)\right],\end{equation}
where the two-level form factor $b(k)$ is the Fourier transform of the
two-level cluster function $Y(x)=[K(x)]^2$.  For the kernel given by
(1), we obtain
\begin{equation}
4\pi b(k)=(|k|+2\pi) {\rm
coth}[(|k|+2\pi)\pi/\beta]+(|k|-2\pi){\rm coth}
[(|k|-2\pi)\pi/\beta]-2|k|{\rm coth}(|k|\pi/\beta).
\end{equation}
Note that $b(k)$ reduces to the exact GUE result in the
limit $\beta\rightarrow 0$.
To an excellent approximation, $b(k)$ contributes negligibly to the
integral in Eq. (4) for $k>2\pi$. We find
\begin{equation}
V_n(s)=[1-\rho]s+\frac{\rho}{\pi^2}\left[Ci(2\pi
s)+1\right]+\frac{1}{\pi^2}\int_{0}^{2\pi s}dk{1-{\rm cos}ks\over k}
\left[\coth\left(\frac{k\pi}{\beta s}\right)-\frac{\beta s}{k\pi}\right],
\end{equation}
where $Ci$ is the cosine integral, and $\rho$ defined as
\begin{equation}
\rho=\rm coth\left(\frac{2\pi^2}{\beta}\right)-\frac{\beta}{2\pi^2}
\end{equation}
goes to 1(0) for $\beta\rightarrow 0(\infty)$.
We can obtain an explicit expression for the number variance by
approximating the factor $[\coth(\frac{k\pi}{\beta s})-\frac{\beta s}{k\pi}]$
under the integral by
$k\pi/3\beta s$ up to $k=3\beta s\rho/\pi$ and by $\rho$ for larger $k$.
The result
\begin{equation}
V_n(s)=[1-\rho]s+\frac{\rho}{\pi^2}\left[1-\frac{{\rm sin}(3\beta\rho
s/\pi)}{3\beta\rho s/\pi}\right]
+\frac{\rho}{\pi^2}\left[\ln\left(\frac{2\pi^2}{3\beta\rho}\right)+
Ci\left(\frac{3\beta\rho s}{\pi}\right)+1\right]
\end{equation}
agrees with the numerical evaluation of the integral up to a small
$\beta$ dependent constant for large $s$ as well as corrections of order
$1/s$. Eq.(8) clearly exhibits the crucial dependence of the number variance
on the parameter $\beta$.
As expected, $\beta=0$ corresponds to the exact GUE result, with a
logarithmic dependence on $s$; increasing $\beta$ corresponds to a
transition towards a Poisson result, which is linear in $s$.  Note
that the variance of an arbitrary linear statistic $f=\sum_{n}f(x_n),$ of
which
$V_n$ is a special case, will depend on the form factor (5) and is
clearly no longer universal,
\begin{equation}
{\rm Var}(f)=\frac{2}{\pi}\int_{0}^{\infty}dk[1-b(k)]
|{\bar {f}}(k)|^2,
\end{equation}
where ${\bar f}(k)$ is the Fourier transform of $f(x)$. This result
generalizes the Dyson-Mehta theorem\cite{Mehta} on the variance of
arbitrary linear statistics and reduces to it when $\beta\rightarrow 0.$
\par\indent
The $\Delta_3$ statistics is a measure of the size of fluctuations of
a given level sequence against a best straight line fit for that level
sequence. If $s(x)$ is the staircase function for a level sequence (in
the variable where the density is uniform) in a given interval, then
one defines a variance $\Delta_3=<{\rm min}_{A,B}[(s(u)-Au-B)^2]>$,
where $<>$ denotes
an average over an ensemble of level sequences. This can
be expressed in terms of the two-level kernel $K(u,v)$ \cite{Mehta}.
For the $q-$Hermite ensemble we obtain the small and large $\beta$
limits explicitly:
\begin{equation}
\Delta_{3}(s,\beta)=\frac{1}{2\pi}\left[\ln(2\pi
s)+\gamma-\frac{5}{4}\right]
+\frac{\beta^{2}s^{2}}{72\pi^{2}}+O\left(\left(\beta
s\right)^{4}\right)
\end{equation}
for $\beta\ll 1/s$, which reduces to the GUE result for $\beta=0$, and
\begin{equation}
{\Delta_{3}(s,\beta)\over s}=\frac{1}{15}\left[\frac{\beta}{2\pi^2}-
\frac{2}{{\rm exp}\left({4\pi^2\over \beta}\right)-1}\right]+C+
O\left({\rm e}^{-\beta s}\right),
\end{equation}
for $\beta\gg 1/s$, which reduces to the Poisson result for
$\beta\rightarrow\infty$. Here $C$ is a numerical constant independent
of $s$ and very weakly dependent on $\beta$. Note that since the ratio
$\Delta_{3}/s$ is
finite, $\Delta_{3}$ always has a linear dependence for large enough
$s$, with a slope that increases with increasing $\beta$, approaching
the Poisson limit for $\beta\rightarrow\infty$. Figure 1 shows the
complete solution (obtained from numerical evaluation of the integrals
involved) for various values of $\beta$. The deviation from the GUE
result as a function of some parameter is qualitatively similar to the
deviations seen numerically in both the transfer matrix
\cite{Muttalib2} and energy eigenvalues \cite{Schrieber} corresponding
to the tight binding Anderson Hamiltonian for disordered conductors, as
well as for the eigenvalues of the evolution operator corresponding to
the Fermi-acceleration model \cite{Jose}.
\par\indent
We find that the statistical properties of the $q-$Hermite (or the
$q-$Laguerre, in the bulk) ensembles are similar to those of a wide
variety of quantum systems, including energy levels of disordered
conductors, studied numerically. Although we are not able to derive
these
ensembles from microscopic Hamiltonians at present,
it is useful to explore specific
spectroscopic signatures of the ensembles so that their relevance to
given physical systems can be tested experimentally. The obvious
problem of looking for evidence of these transitions is the difficulty
to extract a ``stick spectrum'' from experimental data where a large
number of levels are usually lost either in the noise or in unresolved
bands. One way to avoid this problem is to take direct Fourier
transform (FT) of the raw experimental data; the ensemble average of the
square of the FT will show a ``correlation hole'' if the spectrum is
chaotic, i.e. if it has Wigner distribution \cite{Leviandier}. The size
of the
correlation hole is proportional to $[1-b(k)],$ where $b(k)$ is the two
level form factor mentioned before. We plot this function for the
$q-$Hermite ensemble obtained from eq. (5) for various values of $\beta$
in figure 2. The depth of the hole decreases from the Wigner result in
a specific way towards the Poisson limit of no correlation hole ($b(k)=0$).
Thus
e.g. microwave experiments on ensembles of small metallic particles (of
roughly equal size)
\cite{Gorkov} at various disorders might reveal this behavior
(increasing disorder will correspond to increasing $\beta$
\cite{Muttalib}).
Of course one needs sufficiently low temperature and small system size
so that the energy levels are not broadened into a continuum. For a
metallic system, a crude estimate within a simple electron gas model
suggests that for 50 nanometer size particles one might expect to
observe the effect at a temperature in the milikelvin range.
\par\indent
We would like to thank Professors Mike Berry and Peter W\"olfle for
discussions. One of us (K.A.M) would like to thank the Science and
Engineering Research
Council, UK for the award of a Visiting Fellowship, and the Institut f\"ur
Theorie der Kondensierten Materie, Universit\"at Karlsruhe, for kind
hospitality during his stay, supported in part by
Sonderforschungsbereich 195 of Deutsche Forschungsgemeinschaft, when
part of this work was done.
\vskip 1cm
{\bf Figure captions:}
\par\indent
Figure 1: $\Delta_3$ statistics as a function of the length
$s$ of a level sequence
for different values of $\beta$ for the $q-$Hermite model. The $\beta=0$
curve coincides with the GUE result while the $\beta=\infty$ line
coincides with the Poisson result.
\par\indent
Figure 2: The function $[1-b(k)]$ showing the change in the correlation
hole with increasing $\beta$. Interpreting $k$ as time and with the
dashed line replacing an ideal vertical drop at very small time for
finite number of levels $n$, this should correspond to the square of
the Fourier transform of an experimentally obtained spectrum for a
system of small disorderd particles. Increasing $\beta$ will correspond to
increasing disorder \cite{Muttalib}.
\newpage
\noindent
$^{\dag}$Permanent address.


\begin{thebibliography}{99}
\bibitem{Balian}R.Balian, Nuovo Cimento {\bf 57}, 183 (1968).
This is required
whenever the eigenvalues are un-bounded, as in our case, while for
bounded eigenvalues
the confining potential is not necessary.
\end{thebibliography}
\end{document}